\documentclass[aps,showpacs,superscriptaddress,twocolumn]{revtex4-1}
\usepackage{amsmath}
\usepackage{amssymb}
\usepackage{graphicx}
\usepackage{hyperref}
\usepackage{eurosym}
\usepackage{amsfonts}
\usepackage{amssymb}
\usepackage{natbib}
\usepackage[usenames]{color}
\usepackage{color}
\usepackage[T1]{fontenc}
\usepackage[english]{babel}
\usepackage{float}
\usepackage[normalem]{ulem}

\hypersetup{
    colorlinks=true,linkcolor=blue,citecolor=blue,
    filecolor=blue,urlcolor=blue,breaklinks=true
}

\begin{document}

\title{Solution of the $\kappa$-deformed Dirac equation with vector and scalar interactions in the context of spin and pseudospin symmetries}
\author{Claudio F. Farias}
\email{cffarias@gmail.com}
\affiliation{
  Departamento de F\'{i}sica,
  Universidade Federal do Maranh\~{a}o,
  65085-580 S\~{a}o Lu\'{i}s, Maranh\~{a}o, Brazil
}
\author{Edilberto O. Silva}
\email{edilberto.silva@ufma.br}
\affiliation{
  Departamento de F\'{i}sica,
  Universidade Federal do Maranh\~{a}o,
  65085-580 S\~{a}o Lu\'{i}s, Maranh\~{a}o, Brazil
}
\date{\today }

\begin{abstract}
The deformed Dirac equation invariant under the $\kappa$-Poincar\'{e}-Hopf
quantum algebra in the context of minimal and scalar couplings under spin
and pseudospin symmetries limits is considered. The $\kappa$-deformed
Pauli-Dirac Hamiltonian allows us to study effects of quantum deformation in
a class of physical systems, such as an Zeeman-like effect, Aharonov-Bohm
effect and an anomalous-like contribution to the electron magnetic moment,
between others. In our analysis, we consider the motion of an electron in a
uniform magnetic field and interacting with (i) a planar harmonic oscillator
and (ii) a linear potential. We verify that the particular choice of a
linear potential induces a Coulomb-type term in the equation of motion.
Expressions for the energy eigenvalues and wave functions are determined
taking into account both symmetries limits. We verify that the energies and
wave functions of the particle are modified by the deformation parameter as
well as by the element of spin.
\end{abstract}

\maketitle

\section{Introduction}

\label{sec:introduction}

Quantum deformations based on the $\kappa $-Poincar\'{e}-Hopf algebra
constitute an important branch of research that enables us to address
problems in condensed matter and high energy physics through field
equations. These field equations were first presented in Ref. \cite
{PLB.1992.293.344} (see also Refs.
\cite{PLB.1991.264.331,PLB.1993.302.419,PLB.1993.318.613,PLB.1994.329.189,PLB.1994.334.348}), where a new real quantum Poincar\'{e} algebra with standard real
structure, obtained by contraction of $U_{q}\left( O\left(3,2\right) \right)
$. The resulting algebra of this contraction is a standard real Hopf algebra
and depends on a dimension-full parameter $\kappa$ instead of the real
deformation parameter $q$. This algebra is defined by the following commutation relations:
\begin{eqnarray}
&&\left[\Pi_{\nu},\Pi_{\mu}\right]=0, \;\; \mu, \nu = 0,1,2,3, \\
&&\left[M_{i},\Pi_{\mu}\right]=
(1-\delta_{0\mu})i\epsilon_{ijk}\Pi_{k}, \;\; i, j = 1,2,3,\\
&&\left[L_{i},\Pi_{\mu}\right]=i [\Pi_{i}]^{\delta_{0\mu}}
[\delta_{ij}\varepsilon^{-1}
\sinh \left( \varepsilon \Pi_{0}\right)]^{1-\delta_{0\mu}},\\
&&\left[M_{i},M_{j}\right]=i\epsilon_{ijk} M_{k},\\
&&\left[M_{i},L_{j}\right]=i\epsilon_{ijk} L_{k},\\
&&\left[L_{i},L_{j}\right]=-i\epsilon_{ijk}
\left[M_{k}\cosh \left(\varepsilon \Pi_{0}\right)-
\frac{\varepsilon ^{2}}{4}\Pi_{k}\Pi_{l} M_{l}
\right],
\label{eq:algebra}
\end{eqnarray}
where $\varepsilon$ is defined by
\begin{equation}
\varepsilon=\kappa^{-1}=
\lim_{R\rightarrow \infty }(R\ln q),
\end{equation}
with $R$ being the de Sitter curvature, $\Pi_{\mu }=(\Pi_{0},\boldsymbol{\Pi})$
are the $\kappa$-deformed generators for energy and momenta. In the above commutation relations the $M_{i}$, $L_{i}$ represent the spatial rotations and
deformed boosts generators, respectively.
The coalgebra and antipode for the $\kappa$-deformed
Poincar\'{e} algebra was established in Ref. \cite{AoP.1995.243.90}. Since
then, the algebraic structure of the $\kappa$-deformed Poincar\'{e} algebra
has been investigated intensively and havebecome a theoretical field of
increasing interest \cite
{PLB.1994.329.189,PLB.1994.334.348,CQG.2010.27.025012,PLB.2012.711.122,NPB.2001.102-103.161,PLB.2002.529.256,PRD.2011.84.085020,JHEP.2011.1112.080,EPJC.2013.73.2472,PRD.2009.79.045012,EPJC.2006.47.531,EPJC.2008.53.295,PLB.2013.719.467,CQG.2004.21.2179,JHEP.2004.2004.28,PLB.1995.359.339,PLB.1994.339.87,PRD.2013.87.125009,PRD.2012.85.045029,PRD.2009.80.025014}
. Through the field equations from the $\kappa $-Poincar\'{e} algebra ($
\kappa $-Dirac equation \cite
{PLB.1993.302.419,PLB.1993.318.613,EPJC.2003.31.129}), we can study the
physical implications of the quantum deformation parameter $\kappa$ in
relativistic and nonrelativistic quantum systems. In this context, we
highlight the study of relativistic Landau levels \cite{PLB.1994.339.87},
the Aharonov-Bohm effect taking into account spin effects \cite
{PLB.1995.359.339}, the Dirac oscillator \cite
{PLB.2014.731.327,PLB.2014.738.44} and the integer quantum Hall effect \cite
{EPL.2016.116.31002}.

When we want to study the relativistic quantum dynamics of particles with
spin, we must obviously consider the presence of external fields, which
include the vector and scalar fields. The inclusion of vector and scalar
potentials in the Dirac equation reveals interesting properties of
symmetries in nuclear theory. The first contributions in this subject
revealed the existence of $SU(2)$ symmetries, which are known in the
literature as pseudospin and spin symmetries \cite
{AoP.1971.65.352,NPB.1975.98.151}. Some investigations have been made in
this scenario in order to give a meaning to these symmetries. However, it
was only in a work by Ginocchio, that pseudospin symmetry was revealed. He
verified that pseudospin symmetry in nuclei could arise from nucleons moving
in a relativistic mean field, which has an attractive scalar and repulsive
vector potential nearly equal in magnitude \cite{PRL.1997.78.436} (for a
more detailed description see Ref. \cite{PR.2005.414.165}). Spin and
pseudo-spin symmetries in the Dirac equation have been studied under
different aspects in recent years (see Refs. \cite
{JPG.1999.25.811,PRA.2015.92.062137}). Some studies have been developed
taking into account the exact spin and pseudospin symmetries limits to study
the relativistic dynamics of physical systems interacting with a class of
potentials \cite
{CTP.2012.58.807,FBS.2013.54.1839,EPJA.2009.43.73,AMC.2010.216.545,PRA.2012.86.032122,PRC.2012.86.052201,AoP.2015.356.83,AoP.2015.362.196}
.

The present work is proposed to investigate the $\kappa $-deformed Dirac
equation derived in Ref. \cite{PLB.1993.318.613} in the context of minimal
and scalar couplings under spin and pseudospin symmetries limits. The
structure of the paper is as follows. In Sec. \ref{sec:II}, we present the $
\kappa $-deformed Dirac equation with couplings from which we derive the $
\kappa $-deformed Pauli-Dirac equation, by using the usual procedure that
consists of squaring the $\kappa $-deformed Dirac equation. In Sec. \ref
{sec:III}, we consider the equation of Pauli and establish the spin and
pseudospin symmetries limits. As an application, we consider the particle
interacting with an uniform magnetic field in the $z$-direction in two
different physical situations: (i) particle interacting with a harmonic
oscillator and (ii) particle interacting with a linear potential. We obtain
expressions for the energy eigenvalues and wave functions in both limits. In
Sec. \ref{sec:c}, we present our comments and conclusions.

\section{The $\protect\kappa$-deformed Dirac equation with couplings}

\label{sec:II}

We begin with the deformed Dirac equation invariant under the $\kappa$
-deformed Poincar\'{e} quantum algebra \cite
{PLB.1993.302.419,PLB.1993.318.613}
\begin{equation}
\left\{ \left( \gamma _{0}p_{0}-\gamma _{i}p_{i}\right) +\frac{1}{2}
\varepsilon \left[ \gamma _{0}\left( p_{0}^{2}-p_{i}p_{i}\right) -Mp_{0}
\right] \right\} \psi =M\psi .  \label{e1}
\end{equation}
The interactions can be performed through the following prescriptions \cite
{greiner.rqm.wf}:
\begin{eqnarray}
p_{i} &\rightarrow &p_{i}-eA_{i},  \label{coup1} \\
E &\rightarrow &E-\nu \left( r\right) ,  \label{coup2} \\
M &\rightarrow &M+w\left( r\right) .  \label{coup3}
\end{eqnarray}
As mentioned in Ref. \cite{PLB.1993.318.613}, the couplings (\ref{coup1})-(
\ref{coup3}) are quite satisfactory from the point of view of the $\kappa$
-Poincar\'{e}-Hopf algebra. Indeed, there are no operator ordering problems
after the gauging that would require symmetrization in Eq. (\ref{e1}).

As we are interested in a planar dynamics, i.e., when the third directions
of the fields involved are zero, we choose the following representation for
the gamma matrices \cite{NPB.1988.307.909}:
\begin{eqnarray}
\gamma _{0} &=&\sigma _{3},  \label{sigma1} \\
\alpha _{1} &=&\gamma _{0}\gamma _{1}=\sigma _{1},  \label{sigma2} \\
\alpha _{2} &=&\gamma _{0}\gamma _{2}=s\sigma _{2},  \label{sigma3}
\end{eqnarray}
where the parameter $s$, which has a value of twice the spin value, can be
introduced to characterizing the two spin states, with $s=+1$ for spin
\textit{up} and $s=-1$ for spin \textit{down}. In the above representation,
the $\kappa $-deformed Dirac equation including the
interactions can be written as
\begin{eqnarray}
&& \left[ \mathbf{\alpha }\cdot \left( \mathbf{p}-e\mathbf{A}\right) +\gamma
_{0}\left( M+w\left( r\right) \right) \right] \psi -\left[ E-\nu \left(
r\right) \right] \psi  \nonumber \\
&&+\frac{\varepsilon}{2}\left[ es\left( \mathbf{\sigma }\cdot \mathbf{B}
\right) \psi +\gamma _{0}\left( \left( \mathbf{\alpha }\cdot \mathbf{p}
\right) w\left( r\right) \right) \right] \psi  \nonumber \\
&& +\frac{\varepsilon }{2}\left[M\gamma _{0}\left( \mathbf{\ \alpha }\cdot
\mathbf{p}\right)+ \gamma _{0}w\left( r\right) \left( \mathbf{\alpha }\cdot
\mathbf{p}\right) \psi -\gamma _{0}e\left( \mathbf{\alpha }\cdot \mathbf{A}
\right) M\psi \right]  \nonumber \\
&&-\frac{\varepsilon }{2}\left[ \gamma _{0}e\left(\mathbf{\alpha}\cdot
\mathbf{A}\right) w\left( r\right) \psi \right] =0.  \label{d1}
\end{eqnarray}
Let us now determine the Dirac equation in its quadratic form. This can be
accomplished by applying the matrix operator \cite
{AoP.1996.251.45,EPJC.74.3187.2014}
\begin{align}
& \mathbf{\alpha}\cdot \left(\mathbf{p}-e\mathbf{A}\right) +\gamma
_{0}\left( M+w\left(r\right) \right)+ E-\nu \left( r\right) \nonumber
\\
&+\frac{\varepsilon}{2}\left[\gamma _{0}w\left( r\right) \left( \mathbf{
	\alpha }\cdot \mathbf{p}\right)+ es\left(
\mathbf{\sigma}\cdot \mathbf{B}\right) +\gamma_{0}\left( \left( \mathbf{
	\alpha }\cdot \mathbf{p}\right) w\left( r\right) \right) \right]  \nonumber
\\
& +\frac{\varepsilon }{2}\left[M\gamma _{0}\left( \mathbf{\alpha }\cdot \mathbf{p}
\right)  -\gamma _{0}e\left( \mathbf{\alpha }\cdot
\mathbf{A}\right) M-\gamma _{0}e\left( \mathbf{\alpha }\cdot \mathbf{A}
\right) w\left( r\right) \right]
\end{align}
in Eq. (\ref{d1}). The result is the $\kappa $-deformed Dirac-Pauli equation
\begin{align}
& \left( \mathbf{p}-e\mathbf{A}\right) ^{2}\psi +\mathbf{\alpha }\cdot
\left[ \mathbf{p}\nu \left( r\right) \right] \psi -\gamma _{0}\mathbf{\alpha
}\cdot \left[ \mathbf{p}w\left( r\right) \right] \psi \nonumber\\
&+\left[ M+w\left(
r\right) \right] ^{2}\psi-\left[ E-\nu \left( r\right) \right]
^{2}\psi -es\sigma _{z}B\psi \nonumber\\
&-\frac{\varepsilon }{2}\left\{ \gamma _{0}\left[ \mathbf{p}^{2}w\left(
r\right) \right] +\gamma _{0}\left[ \left( \mathbf{\alpha }\cdot \mathbf{p}
\right) w\left( r\right) \right] \left[ \left( \mathbf{\alpha }\cdot \mathbf{
	\ p}\right) \right] \right\} \psi
\nonumber \\
& +\frac{\varepsilon }{2}\left\{ 2is\gamma _{0}\left[ \mathbf{\sigma }\cdot
\left[ \left( \mathbf{p}w\left( r\right) \right) \times \mathbf{p}\right]
-2ise\mathbf{\sigma }\cdot \left( \mathbf{p}w\left( r\right) \right) \times
\mathbf{A}\right] \right\} \psi  \nonumber \\
& +\frac{\varepsilon }{2}\left\{ e\gamma _{0}\left[ \left( \mathbf{\alpha }
\cdot \mathbf{p}\right) w\left( r\right) \right] \left[ \left( \mathbf{\
	\alpha }\cdot \mathbf{A}\right) \right] -w\left( r\right) \left[ \left(
\mathbf{\alpha }\cdot \mathbf{p}\right) w\left( r\right) \right] \right\}
\psi  \nonumber \\
& +\frac{\varepsilon }{2}\left\{ 2MesB+2w\left( r\right) esB-M\left[ \left(
\mathbf{\alpha }\cdot \mathbf{p}\right) w\left( r\right) \right] \right\}
\psi  \nonumber \\
& +\frac{\varepsilon }{2}\left\{ M\gamma _{0}\left[ \left( \mathbf{\alpha }
\cdot \mathbf{p}\right) \nu \left( r\right) \right] +\gamma _{0}w\left(
r\right) \left[ \left( \mathbf{\alpha }\cdot \mathbf{p}\right) \nu \left(
r\right) \right] \right\} \psi =0.  \label{dirackp}
\end{align}
It can be easily verified in Eq. (\ref{dirackp}) that after considering $\varepsilon=0$, the resulting equation is the one well known in the literature
(see, for example, Ref. \cite{EPJC.2015.75.321}). In order to apply the Eq. (\ref{dirackp}) to some physical system, we need to
choose a representation for the vector potential $\mathbf{A}$ and the scalar
potentials $w\left( r\right) $ and $\nu \left( r\right) $. For certain
particular choices of these quantities, we can study the physical
implications of quantum deformation on the properties of various physical
systems of interest.

For the field configuration, we consider a constant magnetic field along the
$z$-direction (in cylindrical coordinates), $\mathbf{B}=B\mathbf{\hat{z}}$,
which is obtained from the vector potential (in the Landau gauge) \cite
{Book.1981.Landau},
\begin{equation}
\mathbf{A}=\frac{Br}{2}\mathbf{\hat{\varphi}}.  \label{Al}
\end{equation}
In this configuration, Eq. (\ref{dirackp}) reads
\begin{equation}
X+\frac{\varepsilon }{2}Y=0,  \label{diracsp}
\end{equation}
with
\begin{eqnarray}
X &=&-\frac{\partial^{2}\psi}{\partial r^{2}}-\frac{1}{r}\frac{\partial \psi
}{\partial r}-\frac{1}{r^{2}}\frac{\partial^{2}\psi}{\partial \varphi^{2}}
+ieB \frac{\partial \psi}{\partial \varphi} +\frac{1}{4}e^{2}B^{2}r^{2}\psi \nonumber \\
&+&
\left[ M+w\left( r\right) \right]^{2}\psi   -\left[ E-\nu \left( r\right) \right]^{2}\psi-es\sigma_{z}B\psi \nonumber \\
&+&i\left[
\frac{\partial w\left(r\right)}{\partial r} \right] \gamma_{0}\alpha_{r}
\psi-i\left[\frac{\partial \nu \left(r\right)}{\partial r}\right]
\alpha_{r}\psi ,
\end{eqnarray}
and
\begin{eqnarray}
Y&=&\gamma_{0}\left[\frac{\partial^{2}w\left(r\right)}{\partial r^{2}}+
\frac{1}{r}\frac{\partial w\left(r\right)}{\partial r}\right]
\psi-\gamma_{0} \left[ \frac{\partial w\left(r\right)}{\partial r}\right]
\frac{\partial \psi}{\partial r} \nonumber \\
&-&is\left[ \frac{1}{r}\frac{\partial
	w\left(r\right)}{\partial r}\right] \frac{\partial \psi}{\partial \varphi}
-2is\left[ \frac{1}{r}\frac{\partial w\left( r\right)}{\partial r}\right]
\frac{\partial \psi}{\partial \varphi} \nonumber \\
&-&iw\left( r\right) \gamma_{r}\left[
\frac{\partial \nu \left(r\right)}{\partial r}\right] \psi +iw\left(
r\right) \alpha_{r}\left[ \frac{\partial w\left(r\right)}{ \partial r}\right]
\psi  \nonumber \\
&+&es\left[\frac{\partial w\left(r\right)}{\partial r}\right] \frac{Br}{2}
\psi -2es\left[\frac{\partial w\left(r\right)}{\partial r}\right] \frac{Br}{2
}\psi \nonumber \\
&-&iM\gamma_{r}\left[\frac{\partial \nu \left(r\right)}{\partial r}
\right]\psi+iM\alpha_{r}\left[\frac{\partial w\left(r\right)}{\partial r}\right] \psi
\nonumber \\
&+&2MesB\psi+2w\left(r\right) esB\psi
\end{eqnarray}
where the matrices (\ref{sigma1})-(\ref{sigma3}) are now given in
cylindrical coordinates, $\gamma_{r}=i\sigma_{\varphi}$, $
\gamma_{\varphi}=-is\sigma_{r}$, with \cite{EPJC.2015.75.321}
\begin{eqnarray}
\alpha_{r} &=&\gamma_{0}\gamma_{r}=\left(
\begin{array}{cc}
0 & e^{-is\varphi} \\
e^{is\varphi} & 0
\end{array}
\right),  \label{alphar} \\
\alpha_{\varphi} &=&\gamma_{0}\gamma_{\varphi}=\left(
\begin{array}{cc}
0 & -ie^{-is\varphi} \\
ie^{is\varphi} & 0
\end{array}
\right),  \label{alphaphi} \\
\gamma_{0} &=&\sigma_{z}=\left(
\begin{array}{cc}
1 & 0 \\
0 & -1
\end{array}
\right).  \label{gammaz}
\end{eqnarray}
We will attribute expressions to functions $\nu \left(r\right)$ and $
w\left(r\right) $ in Eq. (\ref{diracsp}) in the next section, when we treat
analysis of spin and pseudo-spin symmetries. We will argue after that only
some particular choices for these functions will lead to a differential
equation that admits an exact solution.

\section{Symmetries limits}
\label{sec:III}

To implement the spin and pseudospin symmetries limits, we make in Eq. (\ref{diracsp}) the requirement that $w\left( r\right)=\pm \nu \left(r \right)$, where the plus(minus) signal refers to spin(pseudo-spin) symmetry, respectively \cite{PRL.1997.78.436}. Next, by using $\psi=\left(\psi_{+},\psi_{-}\right)^{T}$, the first and second lines in Eq. (\ref{diracsp}) can be written in a simple form, which allows us to solve them
separately. Furthermore, as mentioned above, we need to choose a
representation for the radial function $\nu \left(r \right)$. We give a representation in terms of cylindrically symmetric scalar potentials which lead to results well-known in the literature.

\subsection{Particle interacting with a harmonic oscillator}
\label{sec:A}

Because of applications to various physical systems, we consider the
potential of a harmonic oscillator, $\nu \left( r\right) =\pm w\left(
r\right) =ar^{2}$, where $a$ is a constant. For the Eq. (\ref{diracsp}), one
can check that the spin symmetry is sufficient to decouple the radial
equation that comes from the upper spinor component while the pseudo-spin
symmetry decouples the radial equation that comes from the lower component
of the spinor. Thus, by adopting solutions of the form
\begin{equation}
\psi _{\pm }=\left(
\begin{array}{c}
\sum_{m}f_{+}\left( r\right) e^{im\varphi } \\
i\sum_{m}f_{-}\left( r\right) e^{i\left( m+s\right) \varphi }
\end{array}
\right),  \label{ansatz}
\end{equation}
we arrive at radial equations
\begin{eqnarray}
\frac{d^{2}f_{\pm }(r)}{dr^{2}} &&+\left( \frac{1}{r}+\varepsilon ar\right)
\frac{df_{\pm }(r)}{dr}-\frac{\left( m^{\pm }\right) ^{2}}{r^{2}}f_{\pm
}(r)  \nonumber \\
&&-\left( \omega ^{\pm }\right) ^{2}r^{2}f_{+}(r)+k^{\pm }f_{+}(r)=0,
\label{K1}
\end{eqnarray}
where $k^{+}=E^{2}-M^{2}+2\left(m+s\right) \omega-\varepsilon \left(
2a+3sma+2M\omega s\right) $, $k^{-}=E^{2}-M^{2}+2\omega \left(m+s\right)
+2\omega s-\varepsilon \left[2a-3sa\left( m+s\right)+2M\omega s\right]$, $
\left[\omega^{+}\right]^{2}=\omega^{2}+2\left( M+E\right) a-\varepsilon
\omega sa$, $\left[ \omega^{-}\right]^{2}=\omega^{2}+2\left( E-M\right)
a-\varepsilon \omega sa$, $\omega =eB/2$, $m^{+}=m$ and $m^{-}=m+s$. It is
convenient to write Eq. (\ref{K1}) in a known canonical form. This can be
accomplished using $f_{\pm}(r)$ as
\begin{equation}
f_{\pm}(\rho )=e^{-\frac{1}{2}\left(\kappa^{\pm }+1\right) \rho }\rho^{\frac{
		1}{2}\left\vert m^{\pm}\right\vert}F_{\pm}\left(\rho \right) ,\;\;\rho
=\omega^{\pm }r^{2}  \label{fA}
\end{equation}
where $\kappa ^{\pm }=\varepsilon a/2\omega ^{\pm }$, which leads to the
equation
\begin{eqnarray}
\rho \frac{d^{2}F_{\pm}}{d\rho ^{2}} &+&\left( 1+\left\vert m^{\pm}\right\vert -\rho \right) \frac{dF_{\pm}}{d\rho} \nonumber \\
&-&\left[ \frac{1}{2}\left(1+\left\vert m^{\pm}\right\vert +\kappa ^{\pm }\right) -\frac{k^{\pm }}{4\omega^{+}}\right] F_{\pm }=0.  \label{Kummer}
\end{eqnarray}
Equation (\ref{Kummer}) is of the confluent hypergeometric equation type and
its solution is given in terms of the Kummer functions. In this manner, the
general solution for Eq. (\ref{K1}) is given by \cite{Book.2010.NIST}
\begin{align}
&f_{\pm }(\rho) =\mathit{c}_{1}\,e^{-\frac{1}{2}\left(1+\kappa^{\pm
	}\right) \rho}\rho^{\frac{1}{2}\left\vert m^{\pm}\right\vert} \nonumber \\
&\times\,{\mathrm{M}
	\left(\frac{1}{2}\left(1+\left\vert m^{\pm }\right\vert +\kappa^{\pm}\right)-
	\frac{k^{\pm}}{4\omega^{\pm}} ,1+\left\vert m^{\pm}\right\vert ,\rho \right)}
\nonumber \\
&+\,\mathit{c}_{2}\,{e^{-\frac{1}{2}\left( 1+\kappa^{\pm}\right) \rho
	}\rho^{- \frac{1}{2}\left\vert m^{\pm }\right\vert}}\nonumber \\
&\times\, {{\mathrm{M}\left(\frac{
			1}{2}\left( 1-\left\vert m^{\pm }\right\vert +\kappa^{\pm}\right) -\frac{
			k^{\pm}}{4\omega^{\pm}} ,1-\left\vert m^{\pm}\right\vert ,\rho \right),}}
\end{align}
where $\mathrm{M}$\ are the Kummer functions. In particular, when $\left(
1+\left\vert m^{\pm }\right\vert +\kappa ^{\pm }\right) /2-k^{\pm
}/4\omega^{\pm }=-n$, with $n=0,1,2,...$, the function $\mathrm{M}$ becomes
a polynomial in $\rho $ of degree not exceeding $n$. From this condition, we
extract the energies for the spin and pseudospin symmetries limits, given
respectively by
\begin{align}
E^{2}&-M^{2}=2\sqrt{\omega ^{2}+2\left( M+E\right) a-\varepsilon \omega sa} \notag \\
& \times \left( 2n+\left\vert m\right\vert+1\right) +\varepsilon \left( 3a+3sma+2\omega Ms\right)\notag \\
&-2\omega\left(
m+s\right),  \label{energy1} \\
E^{2}&-M^{2}=2\sqrt{\omega^{2}+2\left( E-M\right)a+\varepsilon \omega sa}\notag \\
& \times\left( 2n+1+\left\vert m+s\right\vert \right)-2\omega \left( m+s\right)-2\omega s
\notag \\
&-\varepsilon \left[ a-3sa\left( m+s\right)+2M \omega s\right].
\label{energy2}
\end{align}

\begin{figure}[!h!]
	\centering
	\includegraphics[width=0.45\textwidth]{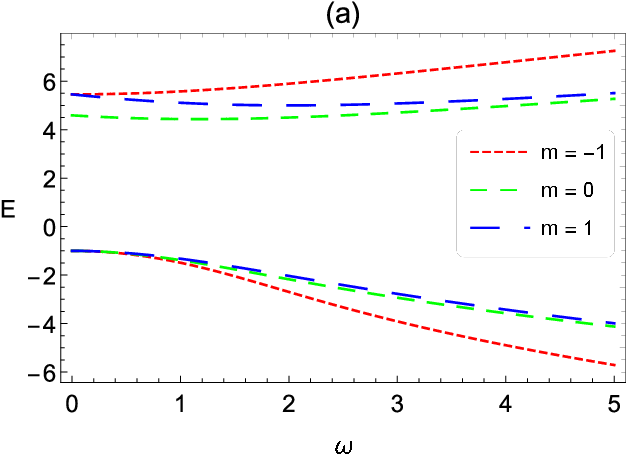}\qquad
	\includegraphics[width=0.45\textwidth]{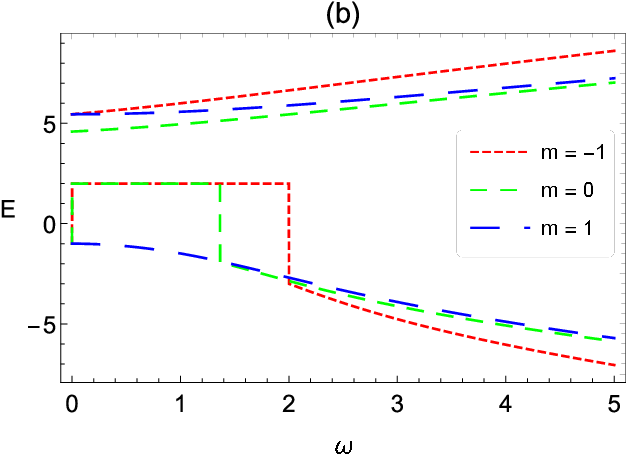}
	\caption{ (Color online) Illustration of the energy eigenvalues in the spin
		symmetry limit (Eq. (\protect\ref{energy1})) as a function of the parameter $
		\protect\omega$ for (a) $s=1$ and (b) $s=-1$. We use units such as $M=1$, $n=1$, $a=1$ and
		an upper bound $\protect\varepsilon=10^{-7}$.}
	\label{Plot_Energy_Eq30}
\end{figure}
\begin{figure}[!h!]
	\centering
	\includegraphics[width=0.45\textwidth]{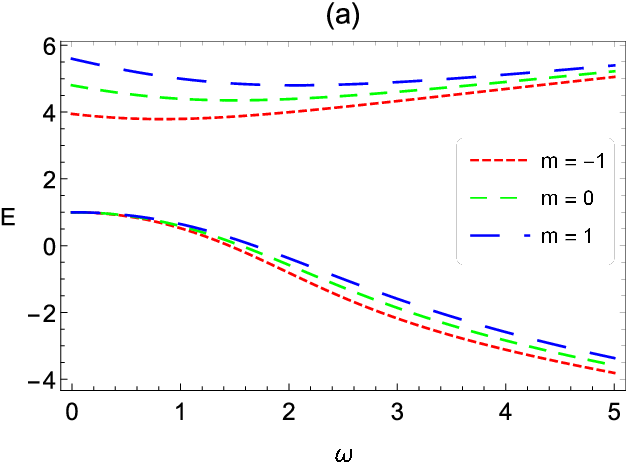}\qquad
	\includegraphics[width=0.45\textwidth]{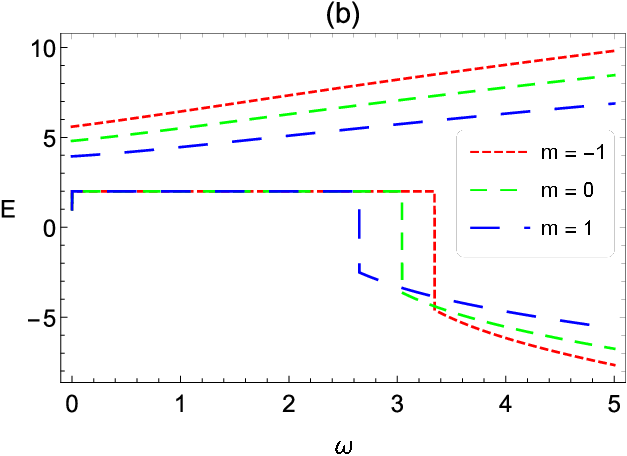}
	\caption{ (Color online) Illustration of the energy eigenvalues in the
		pseudospin symmetry limit (Eq. (\protect\ref{energy2})) as a function of the
		parameter $\protect\omega$ for (a) $s=1$ and (b) $s=-1$. We use units such as $M=1$, $n=1$, $a=1$ and
		an upper bound $\protect\varepsilon=10^{-7}$.}
	\label{Plot_Energy_Eq31}
\end{figure}
Equations (\ref{energy1}) and (\ref{energy2}) are, respectively, the particle
and antiparticle energies in the context of quantum deformation and they can be
read as a relativistic generalization of the Landau levels. It must be
emphasized that, since $\varepsilon$ and $a$ are positive, the quantum
deformation affects the separation of the energy levels of the system. This
feature, however, should not depend on the value of the spin projection
parameter $s$. Figure \ref{Plot_Energy_Eq30} shows the energy profile (\ref{energy1}) as a function of the frequency $\omega$ for some values of the quantum number $m$. In Fig. \ref{Plot_Energy_Eq30}(a), we plotted for $s = 1$ and in the Fig. \ref{Plot_Energy_Eq30}(b) for $s = -1$. In this analysis as well as for the others we use $\varepsilon = 10^{-7}$ \cite{PLB.2013.719.467,EPL.2016.116.31002}. For this value, the effects of quantum deformation become more evident. We clearly see that both particle and antiparticle belong to the same energy spectrum. However, in Fig. \ref{Plot_Energy_Eq30}(b), we find that the antiparticle energy is not defined in the frequency ranges $0<\omega<1.37$ (for $m=0$) and $0<\omega<2.0$ (for $m =-1$). These same characteristics are also present in the energy profile of the Eq. (\ref{energy2}) as shown in Fig. \ref{Plot_Energy_Eq31}. However, in Fig. \ref{Plot_Energy_Eq31}(b), the energies are not defined in the frequency ranges $0<\omega<3.35$ (for $m=-1$), $0<\omega<3.1$ (for $m =-1$) and $0<\omega<2.65$ (for $m =1$). The appearance of such regions characterizing the absence of energy eigenvalues is due to the quantum deformation effects present in the model. In fact, when $\varepsilon=0$, we obtain the energy eigenvalues of
Ref. \cite{EPJC.2015.75.321} (after removing the magnetic flux parameter),
which shows the consistency of the model in question. In particular, when
$a$ and $\varepsilon $ are null, we obtain
\begin{align}
E^{2}-M^{2} &=2\omega \left[ 2n+1+\left\vert m\right\vert -m-s\right] , \\
E^{2}-M^{2} &=2\omega \left[ 2n+1+\left\vert m+s\right\vert -\left(
m+s\right) -s\right],
\end{align}
which are the usual relativistic Landau levels with the inclusion of the
element of spin.

\subsection{Particle interacting with a linear potential}

\label{sec:B}

Let us consider the case where the particle interacts with a linear
potential, $br$. In this case, we make $\nu \left( r\right) =\pm w\left(
r\right) =br$ (where $b$ is a constant) in Eq. (\ref{diracsp}) to the limits
of spin and pseudo-spin symmetries and proceed as before. The resulting
equation is given by
\begin{eqnarray}
&&\left[\frac{d^{2} }{dr^{2}}+\left( \frac{1}{r}+\frac{ \varepsilon b}{2}
\right) \frac{d}{dr}-\frac{\left(m^{\pm }\right) ^{2}}{r^{2}}-\omega
^{2}r^{2}-\mu ^{\pm }r \right]f_{\pm
}\left(r\right)
\nonumber\\
&&-\left(\frac{k^{\pm }}{r} -l^{\pm }\right)f_{\pm
}\left(r\right)=0,  \label{f}
\end{eqnarray}
with $m^{+}=m$, $m^{-}=m+s$, $\mu ^{+}=2\left( E+M\right) b+3\varepsilon
\omega bs/2$, $\mu ^{-}=2\left( E-M\right) b-3\varepsilon \omega bs/2$, $
k^{+}=\varepsilon b\left( 1+3sm\right) /2$, $k^{-}=\varepsilon b\left[
1-3s\left( m+s\right) \right] /2$, $l^{+}=E^{2}-M^{2}+2\omega \left(
m+s\right) -2M\varepsilon \omega s$ and $l^{-}=E^{2}-M^{2}+2\omega \left(
m+s\right) +2\omega s\left( 1-\varepsilon M\right) $. In Eq. (\ref{f}), the $
+(-)$ sing refer to spin and pseudo-spin symmetries, respectively. By
performing the variable change, $x=\sqrt{\omega }r$, Eq. (\ref{f}) assumes
the form
\begin{eqnarray}
&&\left[\frac{d^{2}}{dx^{2}}+\left( \frac{1}{x}+\frac{ \kappa }{2}\right)
\frac{d}{dx}-\frac{\left( m^{\pm}\right) ^{2}}{x^{2}}-x^{2} -a_{L}^{\pm }x\right]f_{\pm}\left(
x\right)
\nonumber
\\
&&-\left(
\frac{a_{C}^{\pm }}{x} -\frac{l^{\pm }}{\omega }\right)f_{\pm}\left(
x\right)=0,  \label{fb}
\end{eqnarray}
where we have defined the parameters $\kappa =\varepsilon b/\sqrt{\omega }$,
$a_{L}^{\pm }=\mu ^{\pm }/\omega \sqrt{\omega }$ e $a_{C}^{\pm }=k^{\pm }/
\sqrt{\omega }$. Note that the choice $\nu \left( r\right) =br$ induces a
Coulomb-like interaction in the resulting deformed sector of the eigenvalue
equation. The origin of this Coulomb potential is due purely to the quantum
deformation and boundary symmetries involved.

Equation (\ref{fb}) is of the Heun equation type, which is a homogeneous,
linear, second-order, differential equation defined in the complex plane.
This equation can be put into its canonical form using the solution
\begin{equation}
f_{\pm }\left( x\right) ={x}^{\left\vert m^{\pm }\right\vert }e{^{-\frac{1}{2}{x}^{2}}}e{^{-\frac{1}{2}\left( a_{L}^{\pm }+\frac{1}{2}\kappa \right) x}y}
_{\pm }\left( x\right) ,  \label{canonical}
\end{equation}
where ${y}_{\pm }$ satisfies the biconfluent Heun differential equation
\begin{align} 
{y}_{\pm}^{^{\prime \prime}}&+\left(\frac{\alpha^{\pm }+1}{x}-2{x}-\beta^{\pm}\right) {y}_{\pm}^{\prime}+\left(\gamma^{\pm}-\alpha^{\pm}-{2}\right) y_{\pm }\notag \\
&-\frac{1}{2x} \left[ {\beta^{\pm}} \left(\alpha^{\pm}+{1}\right)+\delta^{\pm}\right] y_{\pm}=0, \label{heunb}
\end{align}
with $\alpha ^{\pm }=2{\left\vert m^{\pm}\right\vert }$, $\beta ^{\pm}=a_{L}^{\pm}$, $\gamma^{\pm}=\left( \beta ^{\pm }\right) ^{2}/4+l^{\pm}/\omega $ and $~\delta^{\pm}={\kappa /2}+2a_{C}^{\pm}$. Equation (\ref{heunb}) has a regular singularity at $x=0$, and an irregular singularity at
$\infty $ of rank $2$. Usually, the solution of this equation is given in
terms of two linearly independent solutions as
\begin{eqnarray}
y_{\pm }\left( x\right) &=&\mathit{N}\left( \alpha ^{\pm },\beta ^{\pm
},\gamma ^{\pm },\delta ^{\pm };x\right) \nonumber
\\
&+&{x}^{-\alpha ^{\pm }}\mathit{N}
\left( -\alpha ^{\pm },\beta ^{\pm },\gamma ^{\pm },\delta ^{\pm };x\right) ,
\label{heunc}
\end{eqnarray}
where (assuming that $\alpha ^{\pm }$ is not a negative integer)
\begin{equation}
\mathit{N}\left( \alpha ^{\pm },\beta ^{\pm },\gamma ^{\pm },\delta ^{\pm
};x\right) =\sum\limits_{q=0}^{\infty }\frac{{\mathcal{A}}_{q}^{\pm }\left(
	\alpha ^{\pm },\beta ^{\pm },\gamma ^{\pm },\delta ^{\pm }\right) }{\left(
	1+\alpha ^{\pm }\right) _{q}}\frac{x^{q}}{q!}  \label{frb}
\end{equation}
are the Heun functions. After the insertion of this solution into Eq. (\ref
{heunb}), we find ($q\geqslant 0$)
\begin{equation}
\mathcal{A}_{0}=1,  \label{A0}
\end{equation}
\begin{equation}
\mathcal{A}_{1}^{\pm }=\frac{1}{2}\left[ \delta ^{\pm }+\beta ^{\pm }\left(
1+\alpha ^{\pm }\right) \right] ,  \label{A1}
\end{equation}
\begin{eqnarray}
&&\mathcal{A}_{q+2}^{\pm } =\left\{ \left( q+1\right) \beta ^{\pm }+\frac{1}{
	2}\left[ \delta ^{\pm }+\beta ^{\pm }\left( 1+\alpha ^{\pm }\right) \right]
\right\} \mathcal{A}_{q+1}^{\pm }  \nonumber \\
&&-\left( q+1\right) \left( q+1+\alpha ^{\pm }\right) \left[ \gamma ^{\pm
}-\alpha ^{\pm }-2-2q\right] \mathcal{A}_{q}^{\pm },  \label{recur}
\end{eqnarray}
where
\begin{equation}
\left( 1+\alpha ^{\pm }\right) _{q}=\frac{\Gamma \left( q+\alpha ^{\pm
	}+1\right) }{\Gamma \left( \alpha ^{\pm }+1\right) },\;\;q=0,1,2,3,\ldots .
\end{equation}
From the recursion relation (\ref{recur}), the function $N\left( \alpha
^{\pm },\beta ^{\pm },\gamma ^{\pm },\delta ^{\pm };x\right) $ becomes a
polynomial of degree $n$ if and only if the two following conditions are
imposed:
\begin{equation}
\gamma ^{\pm }-\alpha ^{\pm }-2=2n,\;\;n=0,1,2,\ldots   \label{cndA}
\end{equation}
\begin{equation}
\mathcal{A}_{n+1}^{\pm }=0,  \label{cndB}
\end{equation}
where $n$ is a positive integer. In this case, the $\left( n+1\right) $th
coefficient in the series expansion is a polynomial of degree $n$ in $\delta
^{\pm }$. When $\delta ^{\pm }$ is a root of this polynomial, the $\left(
n+1\right) $th and subsequent coefficients cancel and the series truncates,
resulting in a polynomial form of degree $n$ for $N\left( \alpha ^{\pm
},\beta ^{\pm },\gamma ^{\pm },\delta ^{\pm };x\right) $. From condition (\ref{cndA}), we extract the energies at the spin and pseudo symmetries
limit, given respectively by
\begin{eqnarray}
&&E_{nm}^{2}-M^{2} =2\omega _{nm}\left( {n}+{\left\vert m\right\vert +1}
\right) -\frac{b^{2}}{\omega _{nm}^{2}}\left( E_{nm}+M\right) ^{2}  \nonumber
\\
&&-\frac{3b^{2}}{2\omega _{nm}}\left( E_{nm}+M\right) \varepsilon s+2\omega
_{nm}\left[ \varepsilon Ms-\left( m+s\right) \right] ,  \label{Energy+}
\end{eqnarray}
\begin{eqnarray}
&&E_{nm}^{2}-M^{2} =2\omega _{nm}\left( {n}+{\left\vert m+s\right\vert +1}
\right) -\frac{b^{2}}{\omega _{nm}^{2}}\left( E_{nm}-M\right) ^{2}  \nonumber
\\
&&+\frac{3b^{2}}{2\omega _{nm}}\left( E_{nm}-M\right) \varepsilon s-2\omega
_{nm}\left[ s\left( 1-\varepsilon M\right) +\left( m+s\right) \right] .
\label{Energy-}
\end{eqnarray}
The energy of a physical system must be a function involving all the
parameters present in the equation of motion. In Eqs. (\ref{Energy+}) and (\ref{Energy-}) the parameter $a_{C}^{\pm }$ is absent. However, it can be
restored using the condition (\ref{cndB}) from which we obtain a relation
between such parameter and the frequency of the system. For each value of $n$
fixed, we have a self-energy and its corresponding wave function. Let us
consider the solution (\ref{frb}) up to second-order in $x$\ of the
expansion,
\begin{eqnarray}
N\left( \alpha ^{\pm },\beta ^{\pm },\gamma ^{\pm },\delta ^{\pm };x\right) &=&
\frac{\mathcal{A}_{0}}{\left( 1+\alpha ^{\pm }\right) _{0}}+\frac{\mathcal{A}
	_{1}^{\pm }}{\left( 1+\alpha ^{\pm }\right) _{1}}x
\nonumber \\
&+&\frac{\mathcal{A}
	_{2}^{\pm }}{\left( 1+\alpha ^{\pm }\right) _{2}}\frac{x^{2}}{2!}+\ldots .
\label{Heun}
\end{eqnarray}
If we truncate the series in a term of order $x^{n}$, the resulting finite
series is related, via solution (\ref{canonical}), to the energy level $E_{n}$. Thus, the physical quantity that we can associate most closely
with the series (\ref{Heun}) truncated in the term $x^{n}$ is the energy $E_{n}$ of
the particle with the wave function  (\ref{canonical}). In fact, it is necessary that the series (\ref{Heun}) becomes a $n$-degree polynomial for the system to admit bound states. Thus, by using the relation (\ref{recur}) and Eqs. (\ref{A0})-(\ref{A1}), the
coefficient above $\mathcal{A}_{2}^{\pm }$ can be determined. If we want to
truncate solution (\ref{Heun}) in $x$, we must impose that $A_{1}^{\pm }=0$
through the condition (\ref{cndB}); when we truncate in $x^{2}$, we make $
A_{2}^{\pm }=0$, and so on. For each of these cases, we can establish a
appropriate constraints between conditions (\ref{cndA}) and (\ref{cndB}).
Since the Eqs. (\ref{Energy+})-(\ref{Energy-}) are of the relativistic
Landau level type, we prefer to fix the frequency $\omega $ in order to
obtain an expression for the energies corresponding to each value of $n$.
For this, we rewrite the frequency $\omega $ as $\omega ^{\pm }$, with $\pm$
labelling spin and pseudo-spin as above. After obtaining $\omega _{nm}^{\pm}
$ and replace them in Eqs. (\ref{Energy+})-(\ref{Energy-}), we will have
expressions for the energies involving the quantities $\alpha ^{\pm }$, $
\beta ^{\pm }$, $\gamma ^{\pm }$, $\delta ^{\pm }$, which contains the
coupling constant $a_{C}^{\pm }$, the mass of the particle $M$ and the
quantum angular momentum number $m^{\pm }$. Thus, for $A_{1}^{\pm }=0$, it
means that we are investigating the particular solution for $n=0$. In this
case, from Eq. (\ref{A1}), we have
\begin{equation}
\frac{1}{2}\left( \delta ^{\pm }+\beta ^{\pm }\tilde{m}^{\pm }\right) =0,
\label{scdA}
\end{equation}
where $\tilde{m}^{\pm }=1+\alpha ^{\pm }$. Solving (\ref{scdA}) for $\omega$, we get the relations
\begin{eqnarray}
\omega _{0m}^{+} &=&-\frac{2\left( E_{0m}^{+}+M\right) \left( 1+2{\left\vert
		m\right\vert }\right) }{\varepsilon \left[ 3s\left( {\left\vert m\right\vert
	}+m\right) +\frac{3}{2}\left( s+{1}\right) \right] },  \label{En0A} \\
\omega _{0m}^{-} &=&\frac{2\left( E_{0m}^{-}-M\right) \left( 1+2{\left\vert
		m+s\right\vert }\right) }{\varepsilon \left[ 3s\left( {\left\vert
		m+s\right\vert }+m+s\right) +\frac{3}{2}\left( s-{1}\right) \right] }.
\label{En0B}
\end{eqnarray}
Proceeding in a similar way to $\mathcal{A}_{2}^{\pm }=0$, we find the
following third degree polynomial in $\omega _{1m}^{\pm }$ ($n=1$):
\begin{equation}
\mathcal{A}^{\pm }\left[ \omega _{1m}^{\pm }\right] ^{3}+\mathcal{B}^{\pm }
\left[ \omega _{1m}^{\pm }\right] ^{2}+\mathcal{C}^{\pm }\left[ \omega
_{1m}^{\pm }\right] +\mathcal{D}^{\pm }=0,  \label{eqo}
\end{equation}
with
\begin{align*}
	&\mathcal{A}^{+}=2\left( 1+2{\left\vert m\right\vert }\right) , \\
	&\mathcal{A}^{-}=2\left( 1+2{\left\vert m+s\right\vert }\right) , \\
	&\mathcal{B}^{+}=\mathcal{B}^{-}=0, \\
	&\mathcal{C}^{+}=-\frac{3}{2}\varepsilon b^{2}\left( E_{1m}^{+}+M\right) \Big[\left( 2\left\vert m\right\vert +1\right) s+C^{+} \Big], \\
	&\mathcal{C}^{-}=\frac{3}{2}\varepsilon b^{2}\left( E_{1m}^{-}-M\right) \Big[s\left( 2\left\vert m+s\right\vert +1\right)+C^{-} \Big], \\
	&\mathcal{D}^{+}=-b^{2}\left( E_{1m}^{+}+M\right) ^{2}\left( 1+2{\left\vert
		m\right\vert }\right) \left( 3+2{\left\vert m\right\vert }\right) , \\
	&\mathcal{D}^{-}=-b^{2}\left( E_{1m}^{-}-M\right) ^{2}\left( 1+2{\left\vert
		m+s\right\vert }\right) \left( 3+2{\left\vert m+s\right\vert }\right),
\end{align*}
where $C^{+}=2\left(1+\left\vert m\right\vert \right) \left( 2s\left\vert m\right\vert
	+2ms+s+1\right)$ and $C^{-}=2\left(\left\vert m+s\right\vert +1\right) \left( 2s\left( \left\vert 	 m+s\right\vert +m\right) +s+1\right)$. Equation (\ref{eqo}) has only one real root. For each specific frequency, $\omega _{0m}^{\pm}$, $\omega _{1m}^{\pm}$, we can determine the energies $E_{0m}^{\pm}$ and $E_{1m}^{\pm}$ by the following expressions:
\begin{figure}[!t]
	\centering
	\includegraphics[width=0.45\textwidth]{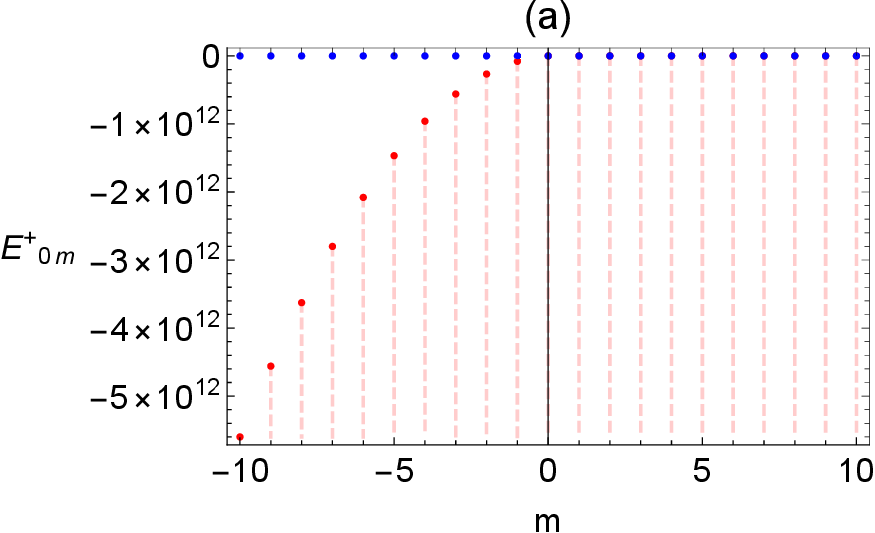}\qquad
	\includegraphics[width=0.45\textwidth]{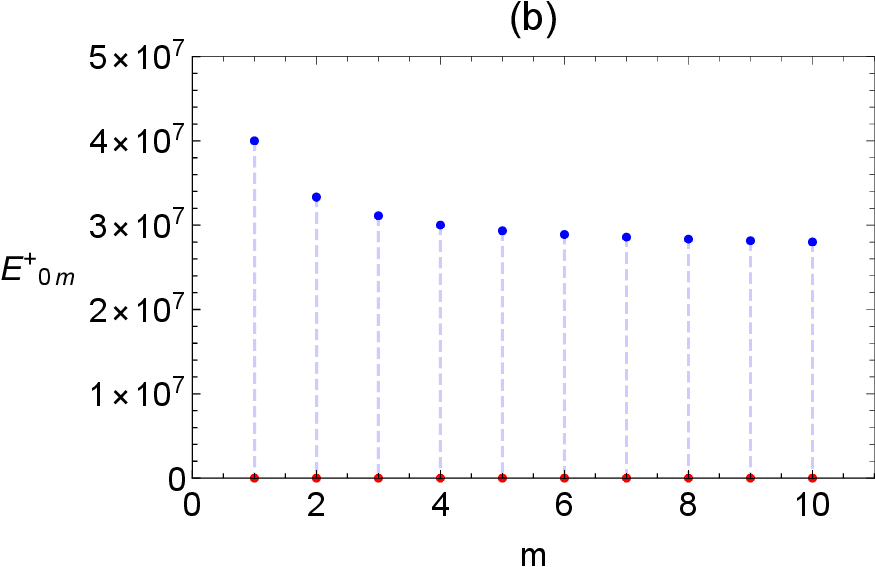}
	\caption{ (Color online) Illustration of the energy eigenvalues in the spin
		symmetry limit (Eq. (\protect\ref{Energy0fi+})) as a function of the parameter $
		\protect\omega $ for (a) $s=1$ and (b) $s=-1$. We use units such as $M=1$, $n=1$, $b=1$ and $\protect\varepsilon=10^{-7}$.}
	\label{Plot_Energy_Eq53}
\end{figure}
\begin{figure}[!t]
	\centering
	\includegraphics[width=0.45\textwidth]{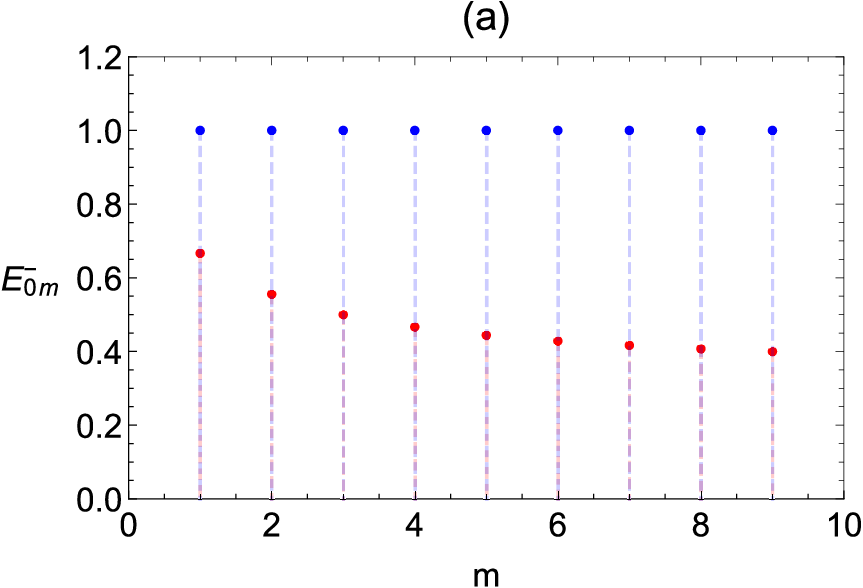}\qquad
	\includegraphics[width=0.45\textwidth]{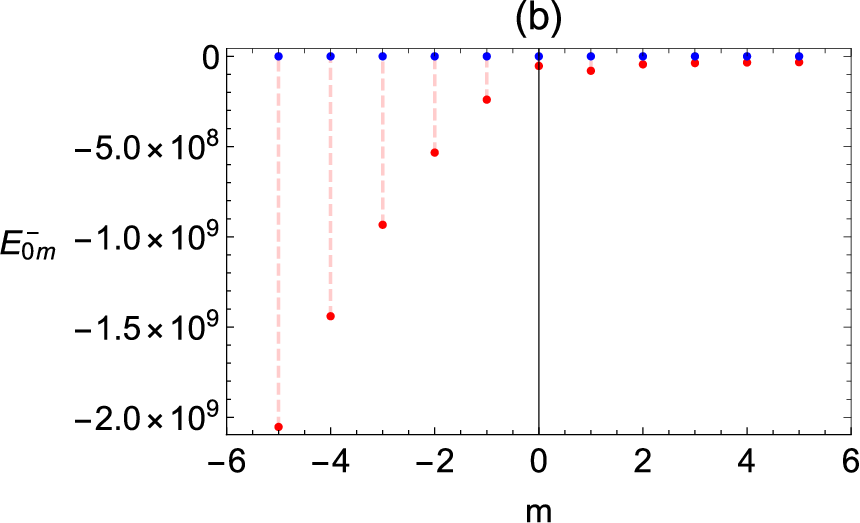}
	\caption{ (Color online) Illustration of the energy eigenvalues in the
		pseudospin symmetry limit (Eq. (\protect\ref{Energy0fi-})) as a function of the
		parameter $\protect\omega $ for (a) $s=1$ and (b) $s=-1$. We use units such as $M=1$, $n=1$, $b=1$ and $\protect\varepsilon=10^{-7}$.}
	\label{Plot_Energy_Eq54}
\end{figure}
\begin{align}
\left( E_{0m}^{+}\right) ^{2}&-M^{2} =2\omega_{0m}^{+}\left( {\left\vert
	m\right\vert +1}\right)\notag \\
& -\frac{b^{2}}{\left( \omega _{0m}^{+}\right) ^{2}}
\left( E_{0m}^{+}+M\right) ^{2} -\frac{3b^{2}}{2\omega _{0m}^{+}}\left( E_{0m}^{+}+M\right) \varepsilon
s,\label{Energy0fi+}
\end{align}
\begin{align}
\left( E_{0m}^{-}\right) ^{2}&-M^{2} =2\omega _{0m}^{-}\left( {\left\vert
	m+s\right\vert +1}\right) \notag \\ &-\frac{b^{2}}{\left( \omega _{0m}^{-}\right) ^{2}}
\left( E_{0m}^{-}-M\right) ^{2}+\frac{3b^{2}}{2\omega _{0m}^{-}}\left( E_{0m}^{-}-M\right) \varepsilon
s\notag \\&-2\omega _{0m}^{-}\left[ s\left( 1-\varepsilon M\right) +\left( m+s\right)
\right]   \label{Energy0fi-}
\end{align}
and
\begin{align}
\left( E_{1m}^{+}\right) ^{2}&-M^{2} =2\omega _{1m}^{+}\left( {\left\vert
	m\right\vert +2}\right) \notag\\ &-\frac{b^{2}}{\left( \omega _{1m}^{+}\right) ^{2}}
\left( E_{1m}^{+}+M\right) ^{2}-\frac{3b^{2}}{2\omega _{1m}^{+}}\left( E_{1m}^{+}+M\right) \varepsilon
s\notag \\&+2\omega _{1m}^{+}\left[ \varepsilon Ms-\left( m+s\right) \right] ,
\label{Energy1fi+}
\end{align}
\begin{align}
\left( E_{1m}^{-}\right) ^{2}&-M^{2} =2\omega _{1m}^{-}\left( {\left\vert
	m+s\right\vert +2}\right) \notag \\ &-\frac{b^{2}}{\left( \omega _{1m}^{-}\right) ^{2}}
\left( E_{1m}^{-}-M\right) ^{2}+\frac{3b^{2}}{2\omega _{1m}^{-}}\left( E_{1m}^{-}-M\right) \varepsilon
s\notag \\&-2\omega _{1m}^{-}\left[ s\left( 1-\varepsilon M\right) +\left( m+s\right)
\right] ,  \label{Energy1fi-}
\end{align}
where $\omega _{1m}^{\pm }$ is given by
\begin{equation}
\omega _{1m}^{\pm }=\sqrt[3]{\frac{\Delta ^{\pm }}{18}}\frac{1}{\mathcal{A}
	^{\pm }}-\sqrt[3]{\frac{2}{3\Delta ^{\pm }}}\mathcal{C}^{\pm },
\end{equation}
with
\begin{equation}
\Delta ^{\pm }=\sqrt{3}\sqrt{27\left[ \mathcal{A}^{\pm }\right] ^{4}\left[
	\mathcal{D}^{\pm }\right] ^{2}+4\left[ \mathcal{A}^{\pm }\right] ^{3}\left[
	\mathcal{C}^{\pm }\right] ^{3}}-9\left[ \mathcal{A}^{\pm }\right] ^{2}
\mathcal{D}^{\pm }.
\end{equation}
Such energies represent the first two energy levels of the system. For simplicity, we perform a numerical analysis only for the case $n=0$. In this way, by studying the Eq. (\ref{Energy0fi+}), we find that for $s = 1$ and $m>0$, the energy of the particle is infinitely degenerate, with respective eigenvalues $E_{0m}^{+}= -1$ and  $E_{0m}^{+}= -0.33$ whereas for $m <0$, only one of the roots is infinitely degenerate with energy $E_{0m}^{+}= -1$ (Fig. \ref{Plot_Energy_Eq53}(a)). On the other hand, when we analyze the Eq. (\ref{Energy0fi+}) for $s=-1$, we verify that the energy spectrum is defined only for $m>0$, and one of the roots is infinitely degenerate with eigenvalue $E_{0m}^{+}= -1$ (Fig. \ref{Plot_Energy_Eq53}(b)). These characteristics are also present in the energies from Eq. (\ref{Energy0fi-}). For $s = 1$, the energies are defined only for $m> 0$ and one of the roots is infinitely degenerate, with eigenvalue $E_{0m}^{-}=1$ (Fig. \ref{Plot_Energy_Eq54}(a)). For $s=-1$ and $m<0$, there is an infinitely degenerate root with eigenvalue $E_{0m}^{-}=1$ while for $m>0$ both roots are infinitely degenerate with respective energies $E_{0m}^{-}=1$ and $E_{0m}^{-}=-2.7 \times10^7$ (Fig. \ref{Plot_Energy_Eq54}(b)).

Finally, to determine the energies corresponding to $n=2,3,4,\ldots $, we must make use of the above recipe. However, as we can see from Eq. (\ref{eqo}), the polynomials of degree $n\geq 3$ resulting from condition (\ref{cndB}), in general, not all roots are
physically acceptable.

\section{Conclusions}

\label{sec:c}

We have studied the relativistic quantum dynamics of a spin-$1/2$ charged
particle with minimal, vector and scalar couplings in the quantum deformed
framework generated by the $\kappa $-Poincar\'{e}-Hopf algebra. The problem
have been formulated using the $\kappa $-deformed Dirac equation in two
dimensions. The $\kappa $-deformed Pauli equation was derived to study the
dynamics of the system taking into account the spin and pseudospin
symmetries limits. For the $\kappa $-deformed Dirac-Pauli equation obtained
(Eq. (\ref{diracsp})), we have argued that only particular choices of radial
function $\nu \left( r\right) $ lead to exactly solvable differential
equations. We have considered the case where the particle interacts with an
uniform magnetic field, a planar harmonic oscillator and a linear potential.
We have verified that the linear potential leads to a Coulomb-type term in
the $\kappa $-deformed sector of the radial equation. The resulting equation
obtained is a Heun-type differential equation. Analytical solutions for both
spin and pseudospin symmetries limits enabled us to obtain expressions for
the energy eigenvalues (through the use of the Eqs. (\ref{cndA}) and (\ref
{cndB})) and wave functions. Because of the limitations imposed by the
condition (\ref{cndB}), we have derived expressions for the energies
corresponding only to $n=0$ (Eqs. (\ref{Energy0fi+})-(\ref{Energy0fi-})) and
$n=1$ (Eqs. (\ref{Energy1fi+})-(\ref{Energy1fi-})). We have shown that the
energy eigenvalues and wave functions are modified by both the spin element $s$ and the deformation parameter $\varepsilon$. We believe that future experiments may provide
some estimate on the magnitude of the deformation parameter within the
context of the model studied.

\section*{Acknowledgments}

EOS acknowledges funding from Conselho Nacional de Desenvolvimento Cient
\'{\i}fico e Tecnol\'{o}gico (CNPq), Grants No. 427214/2016-5 and
No. 303774/2016-9 (PQ), Funda\c{c}\~{a}o de Amparo \`{a} Pesquisa e ao
Desenvolvimento Cient\'{\i}fico e Tecnol\'{o}ico do Maranh\~{a}o (FAPEMA),
Grants No. 01852/14 and No. 01202/16.

\bibliographystyle{apsrev4-1}
\bibliography{dirac-kappa.bbl}

\end{document}